# Transport simulations in hierarchically disordered nanostructures for thermoelectric material design

Laura de Sousa Oliveira[1], Vassillios Vargiamidis[2], and Neophytos Neophytou[3]
School of Engineering, University of Warwick, Coventry, CV4 7AL, UK
[1]L.De-Sousa-Oliveira@warwick.ac.uk, [2]V.Vargiamidis@warwick.ac.uk, [3]N.Neophytou@warwick.ac.uk

*Abstract*—Hierarchically nanostructured materials, where disorder is introduced in various length scales (at the atomic scale, the nanoscale, and the mesoscale) is one of the most promising directions to achieve extremely low thermal conductivities and improve thermoelectric performance. Here we theoretically investigate one such system, a nanocrystalline material with nanopores that are introduced between the crystalline regions. We use the Nonequilibrium Green's Function method for electronic transport and Molecular Dynamics for phonon transport.

## I. INTRODUCTION

The pervasiveness of silicon (Si), not only in its abundance as a natural resource, but in its widespread use in electronics and the corresponding availability of scientific and technological knowledge, make it a desirable material for thermoelectric (TE) applications as well. Thermoelectric performance is quantified by the dimensionless figure of merit $ZT = \sigma S^2 T/(\kappa_e + \kappa_l)$ where $\sigma$ is the electrical conductivity, $S$ is the Seebeck coefficient, $T$ is the operating temperature, $\kappa_e$ is the electronic thermal conductivity, and $\kappa_l$ is the lattice thermal conductivity. The $\sigma S^2$ term is defined as the power factor (*PF*). High bulk thermal conductivity ~130 W m$^{-1}$K$^{-1}$ [2] has, however, limited the use of silicon in TE devices, despite the moderately high power factor (*PF*) — with an upper bound of 6.3 mW m$^{-1}$K$^{-2}$ at 350 K [1]. One of the maximum recorded *ZT* for bulk Si is $ZT = 0.022$, whereas good TE materials have $ZT \sim 1$ (10% of Carnot efficiency), but $ZT \sim 3$ is needed for large scale TE applications. Lowering the thermal conductivity by introducing defects into the material is now the most promising way to decrease the thermal conductivity and improve performance. More specifically, hierarchical nanostructuring, where defects of various length scales are introduced, scatters phonons of various wavelengths, and reduces $\kappa_l$ substantially. However, this is often done at the expense of scattering charge carriers as well. In recent experimental work, however, Lorenzi *et al.* [3] measured a surprising 22 mW m$^{-1}$K$^{-2}$ *PF* for heavily boron (B) doped nanocrystalline silicon films of ~30 nm grain sizes, embedded with nanovoids of a few nanometers in diameter. In previous work, *PF*s up to 15 mW m$^{-1}$K$^{-2}$ have also been achieved for silicon-boron [4]. Furthermore, in nanostructures with dislocation loops it was also demonstrated that the *PF* can increase, in that case by 70% [5]. In those works, the large *PF* was attributed partially to the combined effect of high doping on the electrical conductivity ($\sigma$), and carrier energy filtering by

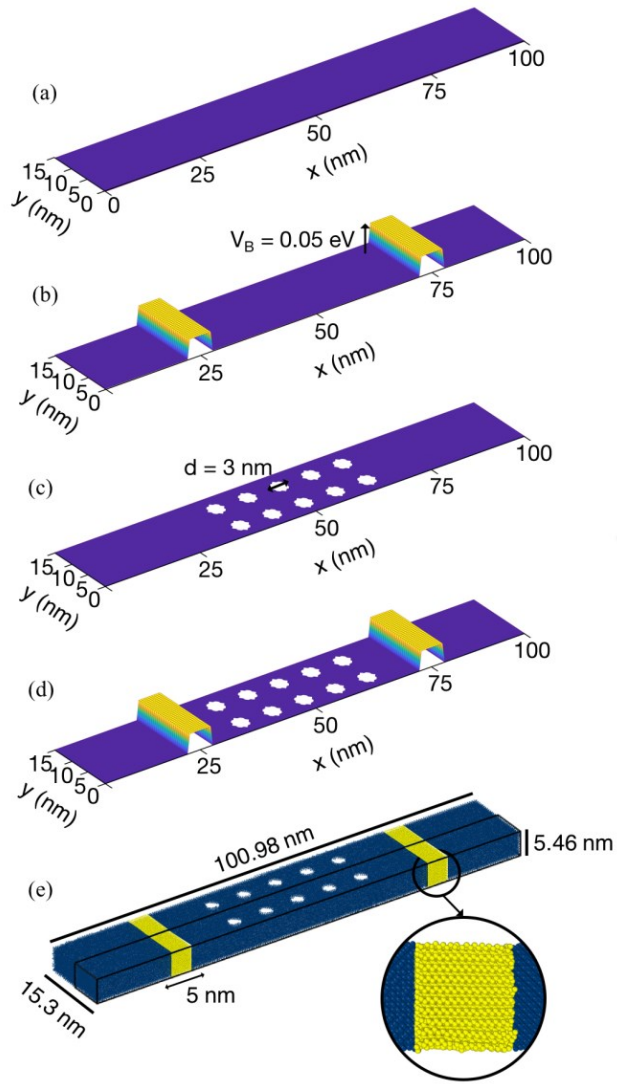

Fig. 1. Two-dimensional channels with embedded SL barriers and/or pores (a)–(d) and equivalent MD geometry. (a) Pristine channel. (b) Channel with means of potential barriers at the grain boundaries on the Seebeck (*S*) coefficient.

In this work we aim to shed some light on two effects: i) Is it possible for the *PF* not to degrade in highly disordered materials, and ii) how much improvement in *ZT* can be achieved in such structures. We use the Nonequilibrium Green's Function (NEGF) approach to evaluate electron transport in two-dimensional (2D) geometries containing supperlatice-type boundaries and pores and Molecular Dynamics (MD) for phonon transport calculations. We find that the combined effect of superlattice barriers and pores under degenerate carrier density conditions can yield a value of *ZT* that is nearly 10× that of pristine Si material.

## II. METHODS

For the electronic calculations, we employ the NEGF method, which can provide flexibility in describing complex geometries [6]. The NEGF 2D simulator takes into account electron-phonon (e-ph) scattering in the self-consistent Born approximation [7]. In this work, we only employ acoustic (elastic) phonon scattering. Four geometries are considered as shown in Fig. 1: a pristine system, a superlattice (SL) with two barriers and no pores, a system with only pores, and a SL with pores. The channel length is 100 nm, and the width 15 nm. The pores have circular shape with a diameter of 3 nm and are arranged in a 5×2 array. The scattering strength of electrons by acoustic phonons is quantified such that the electron mean-free-path is $\lambda = 15$ nm [8] and transport in the channel is diffusive. This is a value comparable to that of silicon [9-11]. For the SLs, the conduction band, $E_c$, is set to 0 eV and the Fermi-level, $E_F$, is aligned with the height of the SL barriers, i.e. $E_F = V_{SL} = 0.05$ eV, for optimal performance, whereas in the pristine channel $E_F = E_c = 0$ eV. Additional code details, including channel calibration, can also be found elsewhere [6]. The *PF*, $GS^2$, where *G* is the conductance and *S* the Seebeck coefficient, is obtained from the following expression, given in terms of the current, *I*:

$$I = G\Delta V + SG\Delta T. \quad (1)$$

Equation (1) is alternately evaluated for $\Delta T = 0$ K and for $\Delta V = 0$ K, with either a small potential or temperature difference, respectively. This approach is validated in [12]. Transmissions are computed as:

$$T = \frac{h}{e^2} \frac{dI}{dV(f_1-f_2)}, \quad (2)$$

where $f_1$ and $f_2$ are the Fermi functions in the left and right contacts. All simulations consider room temperature.

The lattice thermal conductivity calculations are performed within equilibrium MD, with the widely used Green–Kubo [13, 14] approach. The lattice thermal conductivity, $\kappa_l$, of a material is related to fluctuations in its heat-current, *J*, through the Green–Kubo formalism, which can be written as:

$$\kappa_{xx} = \frac{V}{k_B T^2} \int_0^\infty <J_{xx}(t)J_{xx}(t+\tau)>d\tau, \quad (3)$$

where $k_B$ is Boltzmann's constant, *T* the temperature and *V* the volume of the simulated region. In (3), the subscript *xx* denotes the directional thermal conductivity along the *x*-axis, and $<J_{xx}(t)J_{xx}(t+\tau)>$ is the nonnormalized heat-current autocorrelation function (HCACF) in the same direction. This corresponds to the direction of transport in the channel, which was defined to be Si in the [1 0 0] direction in our study. The systems are 184×14×10 supercells of Si, for a diamond cubic 8 atom unit-cell, with periodic conditions in all directions. The geometries closely correspond to the ones used for the electronic calculations, with 1.5 nm radius pores, and 5 nm grain boundaries. The latter is introduced by simply rotating the silicon 45° around the [0 0 1] direction as a very first realization of material change along the transport orientation. Sets of 10 calculations are performed for each of the geometries. It is important to note that the pristine system has been converged for size. The atomic forces of silicon are modeled with the Stillinger–Weber [15] potential, using a set of parameters optimized for thermal transport in silicon based on force matching to density functional theory calculations in the generalized gradient approximation (GGA) [16]. The calculations are performed with the LAMMPS [17] package. After relaxing the atomic structure and lattice parameters using a conjugate gradient algorithm, each system is given a thermal energy equivalent to 300 K (room temperature), also allowing for thermal expansion, by running it for 125 ps, at a 0.5 fs time step, in the isothermal–isobaric (NPT) ensemble. The geometries with grain boundaries are annealed near the contact between the boundaries to allow the system to further relax into a lower energy configuration. The systems are then equilibrated in the microcanonical (NVE) ensemble for another 125 ps.

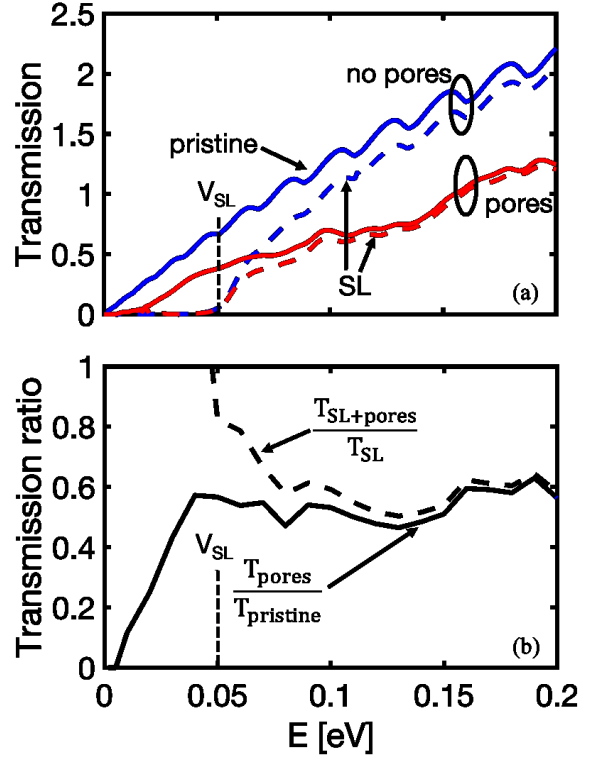

Figure 2. The effects of SL barriers and/or pores in 2D channels on the electronic transmission. (a) Transmission versus carrier energy of a pristine channel in the presence (red solid line) and in the absence (blue solid line) of pores. The dashed lines show the respective transmissions for a channel with SL barriers. (b) Transmission ratio $T_{pores}/T_{pristine}$ and $T_{SL+pores}/T_{SL}$ plotted versus carrier energy.

Finally, a NVE simulation is run for another 10 ns, with a 2 fs time step, during which time the HCACF is recorded.

The electronic results, including the electrical thermal conductance, $K_e$, are combined with the phononic results to estimate the overall effects of SLs and/or pores on the thermoelectric figure of merit, $ZT$. Since the electronic calculations are performed for a 2D system (for numerical efficiency), we can therefore calculate only the electrical thermal conductance. The lattice thermal conductivity is evaluated in 3D, and as a result we get units that can not be combined to compute $ZT$. Due to having different units for thermal transport, $ZT$ for the pristine system is thus computed using experimental results for $\sigma$, $\kappa_e$ and $\kappa_l$. Our results for $\sigma$, $\kappa_e$ and $\kappa_l$ are used to compute the fractional change in these quantities between the pristine channel and other geometries. This allows us to estimate the change in $ZT$ for the various systems based on our calculations. $ZT$ for the geometries with features (SL and pores) are thus computed as:

$$ZT = \frac{\left(\frac{G_{features}}{G_{pristine}}\right)\sigma_{lit}\left(\frac{S_{features}}{S_{pristine}}\right)S_{lit}^2 T}{\left(\frac{K_{e-features}}{K_{e-pristine}}\right)\kappa_{e-lit} + \kappa_{l-lit}\left(\frac{\kappa_{l-features}}{\kappa_{l-pristine}}\right)}, \quad (4)$$

where values from literature [1] are used where the subscript '*lit*' is used in (4). While these values were obtained at 350 K and our calculations performed at 300 K, we find that they represent the best complete set of parameters matching (also the conditions of Lorenzi *et* al. [3]). The values used are included in Table 1.

III. RESULTS

We analyze now the results of our simulations for the geometries shown in Figs. 1(a)–(d). First, we investigate the effects of the SL barriers and/or pores on the electronic transmission of the 2D channels. In Fig. 2(a) we show the transmission, extracted as described in (2), plotted versus the carrier energy. The blue/red lines correspond to the absence/presence of pores in a pristine channel (solid lines) and a channel with SL barriers (dashed lines). It can be seen that the presence of pores degrades the electronic transmission, as expected. In the case of the SL barriers, conduction opens up as soon as the energy reaches the SL barrier height $V_{SL}$. However, after the energy crosses $V_{SL}$ the slope of the transmission is steeper than that of a pristine channel. This indicates that carriers, after passing through the SL barrier, propagate in higher velocity states in the intermediate region. Note also that in a channel with SL barriers the pores are less effective when the energy is above but close to $V_{SL}$, which can be quite important when considering performance optimization in thermoelectric materials. This is reflected in Fig. 2(b) where we show the transmission ratios $T_{pores}/T_{pristine}$ (blue solid line) and $T_{SL+pores}/T_{SL}$ (red dashed line) plotted versus the carrier energy. Here, $T_{pristine}$ and $T_{SL}$ are the transmissions of the pristine channel and of the channel with SL barriers, respectively, while $T_{pores}$ and $T_{SL+pores}$ are the same systems with pores. The presence of pores in an otherwise pristine channel is less effective at higher energies. However, for a channel with SL barriers, the effect of pores is even weaker for energies above but close to $V_{SL}$ compared to the pristine channel. This is quite important, indicating that the electronic conductance in highly disordered structures, which can slow down phonons significantly, can be less affected if they are operated at degenerate conditions, which will help the *PF*.

Table 1 summarizes our results, including electronic conductance, Seebeck coefficient and *PF* of the structures shown in Fig. 1. We compare the values of these quantities for each complex geometry with respect to those of the pristine material. Interestingly, by raising $E_F$ higher in the bands, $G$ increases by 55% despite the introduction of the SL barriers, as discussed above. This is due to higher carrier velocities and degenerate carrier concentration conditions. The Seebeck coefficient decreases, as expected, but overall the *PF* increases in the SL structure by 23% compared to that in the pristine structure. This indicates that the energy filtering provided by potential barriers that cut lower parts of the Fermi distribution is more effective at degenerate conditions, demonstrating that it is possible to have *PF* benefits [4]. However, the improvement of the *PF* originates from the increase of the conductivity rather than from the Seebeck coefficient. The introduction of pores in a pristine channel has a strong degrading effect in the conductance, which is due to additional scattering. Consequently the *PF* decreases by 45% compared to the pristine channel. On the other hand, for a channel with SL barriers and pores in between, the *PF* decreases less, by 35%, as the conductance reduces to values close to those of the pristine channel.

As expected, introducing grain boundaries and pores significantly lowers thermal conductivity. This is largely due to a reduction in the mean free path as a result of phonon scattering at the pore and grain boundary interfaces. The thermal conductivity drops by approximately 5× in the supperlattice geometry compared to the pristine single-crystal silicon. Note that, as a result of annealing, some amorphicity has been introduced near the boundaries of where the different orientation crystals meet. For a mere 4.6 % porosity, the reduction in $\kappa_l$ is 7.1×. Yet, we have only considered low porosities. As porosity increases so does thermal resistance and, at high porosities, the thermal conductivity approaches the amorphous limit for silicon. Should both scattering features (pores and grain boundaries) be independent from each other, one might expect Matthiessen's rule to hold and for the total thermal conductivity of the combined pores and grain boundaries system to be given by $1/\kappa_{SL+pores} = 1/\kappa_{SL} + 1/\kappa_{pores}$, such that $1/\kappa_{SL+pores} \approx 17.2$ Wm$^{-1}$K$^{-1}$. Instead we find the total effect is greater ($\kappa_{SL+pores} = 12.1$ Wm$^{-1}$K$^{-1}$) than expected, and the combined nano-features account for a 17.1× decrease in $\kappa_l$. The results of the Molecular Dynamics simulations can be found in Table 1.

In Table 1 we show simulation results for the $S$, $PF$, $G$, $K_e$ and $\kappa_l$. We remind the reader that $ZT$ for the pristine system is computed from values found in the literature. For $\sigma$ we use the value found in [1] (0.588 mΩ$^{-1}$cm$^{-1}$) for p-type silicon at 350 K with a 8.1×10$^{19}$ cm$^{-3}$ charge carrier concentration. This that best approximates the work of Lorenzi *et* al. [3]. For consistency, we calculate the electronic thermal conductivity from the above value for $\sigma$ using the Wiedemann–Franz law, yielding $\kappa_{e-lit} = 0.5$ Wm$^{-1}$K$^{-1}$. To be consistent, we use the thermal conductivity on the same study, reported as 102 W m$^{-1}$K$^{-1}$ [2], this leaves us is

| | $\sigma$ [kS m$^{-1}$] | $S$ [$\mu$V K$^{-1}$] | $PF$ [mW K$^{-2}$m$^{-1}$] | $\kappa_e$ [W K$^{-1}$m$^{-1}$] | $\kappa_l$ [W K$^{-1}$m$^{-1}$] | $ZT$ |
|---|---|---|---|---|---|---|
| Literature [1] at 350 K | 58.82 | 328 | 6.3 | 0.5* | 101.5* | 0.022 |
| **Geometry** | $G$ [$\mu$S] | $S$ [$\mu$V K$^{-1}$] | $PF$ [mW K$^{-2}$] | $K_e$ [W K$^{-1}$] | $\kappa_l$ [W K$^{-1}$m$^{-1}$] | *Change in ZT* |
| Pristine | 19.0 | 195 | 7.22×10$^{-10}$ | 8.42×10$^{-11}$ | 207.41±11.51 | 1× |
| SL | 29.4 | 174 | 8.87×10$^{-10}$ | 11.9×10$^{-11}$ | 41.99±2.47 | 5.9× |
| Pores | 9.0 | 211 | 4.00×10$^{-10}$ | 3.71×10$^{-11}$ | 29.14±2.48 | 3.9× |
| SL + pores | 17.8 | 162 | 4.71×10$^{-10}$ | 7.79×10$^{-11}$ | 12.10±1.33 | 10.3× |

*Values calculated from [1] as indicated in the text.

Table 1. Thermoelectric coefficients, thermal conductivities, and change in the figure of merit for the structures shown in Fig. 1. The value of ZT on the second row is computed from values in the literature at 350 K. Calculation details are included in the text. Note that the thermal conductivity in MD is higher than experimentally realized due to the choice of the potential used.

$\kappa_{l\text{-}lit}$ = 101.5 W m$^{-1}$K$^{-1}$. These values used in conjuction with the value of $S$ obtained in the same study yield a $ZT$ of 0.022, which is on the upper bound of other reported values [1, 9], for the pristine system. By using our results of $G$, $S$, $K_e$, and $\kappa_l$ to compute fractional changes in the values obtained from the literature, as indicated in (4), we compute the estimated (changes in the) values of $ZT$ based on our calculations. The results can be found in Table 1. The figure of merit shows the highest value of $ZT$ for a channel with SL barriers and pores (10.3× the $ZT$ of the pristine channel), as a result of maximum phonon scattering and enhanced electronic conductivity. This is an order of magnitude difference from the pristine case, and nearly twice the value of the SL-only case. A large increase in the figure of merit, ZT~0.5, has been measured in silicon nanowires, where in that case phonon scattering at the boundaries lowers the lattice thermal conductivity [18].

## IV. CONCLUSIONS

In conclusion, using the fully quantum mechanical None Green's Function method combined with Molecular Dynamics simulations, we investigated the effects of embedded SL barriers and pores on the thermoelectric figure of merit in hierarchically nanostructured geometries. We show that upon nanostructuring, which is the primary cause to larger ZT due to reductions in $\kappa_l$, the $PF$ suffers less if the Fermi level is placed high into the bands, i.e. under highly degenerate conditions. In this way, utilizing the higher conducting states, benefits in the $PF$ can be achieved, or at least the degradation is less severe. We surmise these results could be useful in the design of nanostructured channels for thermoelectric applications.

ACKNOWLEDGMENT

This work has received funding from the European Research Council (ERC) under the European Union's Horizon 2020 Research and Innovation Programme (Grant Agreement No. 678763). The authors thank Samuel Foster and Dhritiman Chakraborty for helpful discussions.